\documentclass[a4paper]{PoS}

\usepackage{amsmath}
\usepackage{amssymb}
\usepackage{bm}
\usepackage{float}
\usepackage{mathbbol}
\usepackage{mathtools}
\usepackage{comment}
\usepackage[document]{ragged2e}
\usepackage{slashed}
\usepackage{euler}
\usepackage{simplewick}
\usepackage{float}
\usepackage{url}
\usepackage[square,numbers]{natbib}
\graphicspath{{./figures}{.}}
\usepackage[dvipsnames]{xcolor}
\usepackage{xcolor}
\usepackage{subfigure}
\usepackage{soul}
\usepackage{tabularx}
\usepackage{tabulary}
\usepackage{enumerate}
\usepackage{comment}
\usepackage{dcolumn}
\usepackage{makeidx}
\usepackage{ifthen}

\usepackage[normalem]{ulem}

\newcommand{\bra}[1]{\left\langle #1\vert\right.}
\newcommand{\ket}[1]{\left.\vert #1\right\rangle}
\newcommand\xcaption[4][] {
    \ifthenelse{\equal{#1}{}}
               {\caption[#3]{\label{#2}#3 #4}}
               {\caption[#1]{\label{#2}#3 #4}}
}

\def\L{\Lambda}

\def\b{\beta}

\def\s{\sigma}
\def\t{\tau}
\def\MSbar{$\overline{MS}$}

\newcommand\eqsl[1]                            
{
\begin{align}
#1
\end{align}
}

\newcommand{\ccny}{
\llap{$^l$}Department of Physics, The City College of New York, New York, NY 10031, USA\\
	}

\newcommand{\julich}{
	\llap{$^d$}Institut f\"{u}r Kernphysik and Institute for Advanced Simulation,  Forschungszentrum J\"{u}lich\\
	54245 J\"{u}lich Germany\\
 }
\newcommand{\lblnsd}{
	\llap{$^c$}Nuclear Science Division, Lawrence Berkeley National Laboratory,	Berkeley, CA 94720, USA\\
	}
\newcommand{\lblnersc}{
	\llap{$^j$}NERSC, Lawrence Berkeley National Laboratory, Berkeley, CA 94720, USA\\
	}
\newcommand{\liverpool}{
	\llap{$^h$}Theoretical Physics Division, Department of Mathematical Sciences, University of Liverpool, Liverpool L69 3BX, UK\\
    }
\newcommand{\llnl}{
	\llap{$^e$}Physics Division,	Lawrence Livermore National Laboratory, Livermore, CA 94550, USA\\
	}
\newcommand{\nvidia}{
   \llap{$^g$}NVIDIA Corporation, 2701 San Tomas Expressway, Santa Clara, CA 95050, USA\\
    }
\newcommand{\rbrc}{
	\llap{$^k$}RIKEN BNL Research Center, Brookhaven National Laboratory,	Upton, NY 11973, USA\\
	}

\newcommand{\unc}{
	\llap{$^a$}Department of Physics and Astronomy, University of North Carolina\\
	Chapel Hill, NC 27516-3255, USA\\
	}
\newcommand{\wm}{
	\llap{$^b$}Department of Physics, The College of William \& Mary,	Williamsburg, VA 23187, USA\\
	}
\newcommand{\ithems}{
\llap{$^f$}Interdisciplinary Theoretical and Mathematical Sciences Program (iTHEMS)\\
RIKEN 2-1 Hirosawa,
Wako, Saitama 351-0198, Japan\\
}
\newcommand{\jlabcomp}{
\llap{$^i$}
Scientific Computing Group, Thomas Jefferson National Accelerator Facility, Newport News, VA 23606, USA\\
}

\title{Short Range Operator Contributions to $0\nu\beta\beta$ decay from LQCD}
\ShortTitle{}
\author{\speaker{H.~Monge-Camacho}$^{a,b,c}$}
\author{
E.~Berkowitz$^{d}$,
D.~Brantley$^{b,c,e}$,
C.C.~Chang$^{f,c}$,
M.A.~Clark$^{g}$,
A.~Gambhir$^{e}$,
N.~Garron$^{h}$,
B.~Jo\'{o}$^{i}$,
T.~Kurth$^{j}$,
A.~Nicholson$^{a}$,
E.~Rinaldi$^{k}$,
B.C.~Tiburzi$^{l}$,
P.~Vranas$^{e}$, 
A. Walker-Loud$^{c}$\\
\unc 
\wm 
\lblnsd 
\julich 
\llnl 
\ithems 
\nvidia 
\liverpool 
\jlabcomp 
\lblnersc 
\rbrc 
\ccny 

E-mail: \email{hjmonge@email.unc.edu}$^*$}

\abstract{
The search for neutrinoless double beta decay of nuclei is believed to be one of the most promising means to search for new physics.
Observation of this very rare nuclear process, which violates Lepton Number conservation, would imply the neutrino sector has a Majorana mass component and may also provide an explanation for the universe matter-antimatter asymmetry of the universe. In the case where a heavy intermediate particle is exchanged in this process, QCD contributions from short range interactions become relevant and the calculation of matrix elements with four-quark operators becomes necessary. In these proceedings we will discuss our current progress in the calculation of these four-quark operators from LQCD.}

\FullConference{The 36th Annual International Symposium on Lattice Field Theory - LATTICE2018\\
		22-28 July, 2018\\
		Michigan State University, East Lansing, Michigan, USA.}

\begin{document}
\justify
\section{Introduction}

One exciting prospect for searches for physics Beyond the Standard Model (BSM) is neutrinoless double beta decay ($0\nu\beta\beta$), a Lepton Number violating process. If observed, $0\nu\beta\beta$ may provide a possible explanation for the observed abundance of matter over anti-matter in the universe, as lepton number violation could be converted to baryon number violation very early in the universe. Its observation would provide evidence of the Majorana nature of the neutrino.

Most of the recent research on $0\nu\beta\beta$ has been focused on the light Marjorana exchange mechanism which is a long-range interaction process.
Short-range interactions, on the other hand, have received much less attention as a natural hierarchy of higher order operators would render them irrelevant. There are good reasons to believe these short-distance operators may be relevant, however.
The light Majorana exchange mechanism requires a helicity flip and is therefore proportional to the light neutrino mass.  If this mass is generated from heavy new physics, it can naturally lead to comparable long and short-range contributions.
For a comparison of the possible effects from long and short range contributions, see Ref.~\cite{Menendez:2017fdf} and for a detailed discussion of the full possibilities of long and short range mechanisms, see Ref.~\cite{Cirigliano:2018yza}. 


Understanding how the long-range Majorana mechanism contributes to $0\nu\beta\beta$ has been discussed extensively in the literature.  For a recent review, see Ref.~\cite{Engel:2016xgb}.
While theoretical challenges remain, it is in principle understood how to compute the nuclear double beta decay rate given the exchange of a light Majorana neutrino, since the coupling of the weak axial current to a nucleon is well understood experimentally and theoretically.
In contrast, in the scenario that short distance 4-quark--2-electron operators are relevant to $0\nu\beta\beta$, the only recourse we have to ultimately determine the strength of these contributions is with lattice QCD as they can not be isolated experimentally.
For this reason, we have begun our effort focused on these prospective short distance contributions.
There are interesting challenges in understanding the long-range contributions as well~\cite{Cirigliano:2018hja} for which lattice QCD can also be applied~\cite{Feng:2018pdq,Detmold:2018zan}.


\section{Short Range Contributions}
\begin{figure}
\centering
\includegraphics[width=0.9\textwidth]{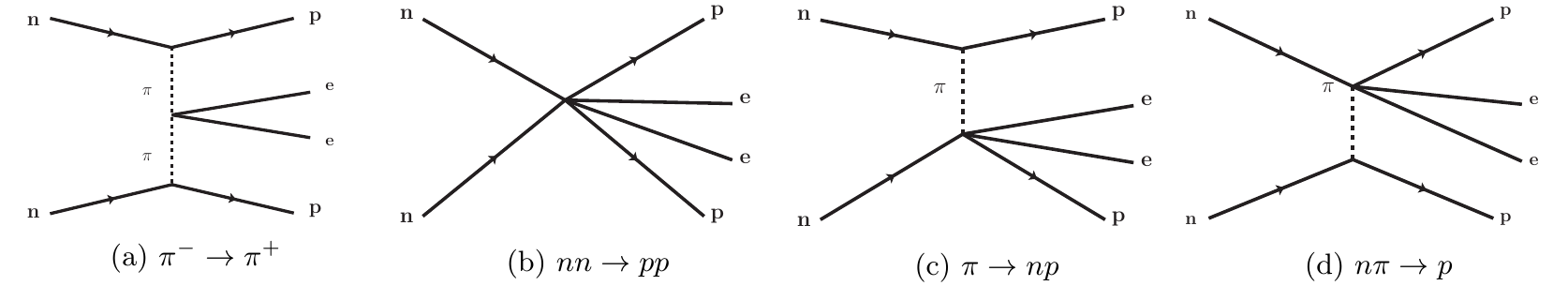}\caption{Diagrams for processes contributing to $0\nu\beta\beta$ through short range interactions.}\label{0nbbDiags}
\end{figure}

$0\nu\beta\beta$ decay violates lepton number conservation by 2 units and thus it is not allowed in the Standard Model (SM). Nevertheless, Lepton number conservation is anomalous in the SM and it might be violated at energies higher than the electroweak scale. To incorporate those effects into the nuclear interactions, a series of Effective Field Theories (EFTs) can be employed, see for example~\cite{Cirigliano:2018yza}. Using the EFT approach, the leading short range
contributions to $0\nu\beta\beta$, which correspond to 9 local four-quark operators, have been identified~\cite{Prezeau:2003xn,Graesser:2016bpz}. These are ilustrated in figure~\ref{0nbbDiags}.

Here we focus on the processes arising from the $0^+ \rightarrow 0^+$ nuclear transitions which are the most relevant for $0\nu\beta\beta$ experimental searches. At leading order in an EFT employing a Weinberg power counting~\cite{Prezeau:2003xn,Graesser:2016bpz}, these are given by the processes shown in figures~\ref{0nbbDiags}(a) and~\ref{0nbbDiags}(b). The matrix element arising from $\pi^-\rightarrow \pi^+$ are the simplest to compute and our Lattice QCD  results~\cite{Nicholson:2016byl,Nicholson:2018mwc} will be presented next. The local $nn \rightarrow pp$ contribution is significantly more challenging to compute due to the challenge of computing two-nucleon systems with LQCD.

The operators that must be considered to compute the $\pi^- \rightarrow \pi^+$ matrix element have been identified in~\cite{Prezeau:2003xn,Graesser:2016bpz}. These operators can be written as three color-unmixed (quarks in a bilinear are contracted together) and two color-mixed (quarks in different bilinears are contracted together) operators. The mixing patern is determined by the chiral properties of these operators~\cite{Graesser:2016bpz,Boyle:2017skn,Donini:1999sf}. The operator basis is given by:
\begin{eqnarray}
\label{eq:Ops}
\mathcal{O}_{1+}^{++}& = &\left(\bar{q}_L \tau^+ \gamma^{\mu}q_L\right)\left[\bar{q}_R \tau^+\gamma_{\mu} q_R \right], \qquad
\mathcal{O}_{1+}^{'++} = \left(\bar{q}_L \tau^+ \gamma^{\mu}q_L\right]\left[\bar{q}_R \tau^+\gamma_{\mu} q_R \right) \ \cr
\mathcal{O}_{2+}^{++}& = &\left(\bar{q}_R \tau^+ q_L\right)\left[\bar{q}_R \tau^+ q_L \right] + L \leftrightarrow  R, \qquad \mathcal{O}_{2+}^{'++} = \left(\bar{q}_R \tau^+ q_L\right]\left[\bar{q}_R \tau^+ q_L \right) + L \leftrightarrow R \  \cr
\mathcal{O}_{3+}^{++}& = &\left(\bar{q}_L \tau^+ \gamma^{\mu}q_L\right)\left[\bar{q}_L \tau^+ \gamma_{\mu} q_L \right] + \left(\bar{q}_R \tau^+ \gamma^{\mu}q_R\right)\left[\bar{q}_R \tau^+ \gamma_{\mu} q_R \right]
\end{eqnarray}
Here, following the Takahashi notation, color contractions are indicated by using () and [] to enclose contracted quarks, for example $\bar{q}_a q_a\bar{q}_b q_b=(\bar{q}q)\left[\bar{q}q \right]$.
In order to calculate the matrix elements, three point functions $C_i^{3\mathrm{pt}}(t,T-t)$ are computed on the lattice. These correspond to correlation functions with the four-quark operator inserted between two pions interpolating fields as follows: 
\begin{equation}\label{eq:C3pt}
C^{3pt}_i(t_i,t_f)=\sum_{\alpha}\sum_{\mathbf{x},\mathbf{y}}e^{-E_\alpha T}\bra{\alpha}\mathcal{\Pi^+}(t_f,\mathbf{x}) \mathcal{O}_i(0,\mathbf{0})\mathcal{\Pi^+}(t_i,\mathbf{y}) \ket{\alpha}
\end{equation}
where $\mathcal{\Pi^+}(t_f,\mathbf{x})=\bar{d}\gamma^5u$ and $\mathcal{\Pi^+}(t_i,\mathbf{y})=\mathcal{\Pi^-}^\dagger(t_i)$ are annihilation and creation operators for a $\pi^+$ and a $\pi^-$ particle respectively. The calculations were performed with $N_f=2+1+1$ HISQ~\cite{Follana:2006rc} gauge field configurations generated by the MILC collaboration~\cite{Bazavov:2010ru,Bazavov:2012xda} and using a Mixed Action with M\"{o}bius Domain Wall valence fermions~\cite{Berkowitz:2017opd}. A more detailed description can be found in~\cite{Nicholson:2016byl,Nicholson:2018mwc}. From $C_i^{3\mathrm{pt}}(t,T-t)$, the following ratio functions $\mathcal{R}_i(t) $ can be obtained and related to the matrix element as follows:
\begin{align}\label{eq:ratio}
\mathcal{R}_i(t) \equiv C_i^{3\mathrm{pt}}(t,T-t)/\left(C_{\pi}(t)C_{\pi}(T-t)\right)=\frac{a^4\bra{\pi}\mathcal{O}^{++}_{i+}\ket{\pi}}{(a^2Z_0^\pi)^2}+\mathcal{R}_{e.s.}(t)
\end{align}
Sample plots for~\ref{eq:ratio} and the extracted matrix element are found in~\cite{Nicholson:2018mwc} figures 2 and 3 respectively. 
After renormalization, to be discussed next, the matrix elements are extrapolated to the continuum, infinite volume and physical pion mass limits~\cite{Nicholson:2018mwc}.
These matrix elements, $O_i$, can then be used to construct $nn\rightarrow ppe^-e^-$ potentials which can be embedded in many body wave functions to compute a $0\nu\beta\beta$ decay rate.
For $i=[1,2]$, the potentials are
\begin{align}\label{eq:2nucPot}
V^{nn\rightarrow pp}_i(|\mathbf{q}|) &=
	-O_i P_{1+}P_{2+}\frac{\partial}{\partial m_\pi^2} V_{1,2}^\pi(|\mathbf{q}|)= -O_i \frac{g_A^2}{4F_\pi^2}
		\t_1^+ \t_2^+
		\frac{\mathbf{\s}_1 \cdot \mathbf{q}\, \mathbf{\s}_2 \cdot \mathbf{q}}{(|\mathbf{q}|^2 + m_\pi^2)^2}\,
\end{align}
where $V_{1,2}^\pi(|\mathbf{q}|) = -\t_1 \cdot \t_2\, \s_1 \cdot \mathbf{q}\, \mathbf{\s}_2 \cdot \mathbf{q} / (|\mathbf{q}|^2 + m_\pi^2)$ and $P_{1,2}^+$ project onto the isospin raising operator for each nucleon (the electrons $\bar{e}e^c$ and the prefactor $\frac{G_F^2}{\L_{\b\b}}$ are not shown).


\smallskip\noindent\textbf{Non-perturbative Renormalization:}
The bare LQCD matrix elements must be renormalized.  It is also useful to convert them to $\overline{MS}$-scheme, the most common in the phenomenological and perturbative QCD literature.
We use the RI/SMOM scheme~\cite{Sturm:2009kb} which is a RI/MOM scheme~\cite{Martinelli:1994ty} with non-exceptional kinematics, with a renormalization scale given by $\mu^2\equiv(p_{in}-p_{out})^2=p_{in}^2=p_{out}^2$.  The renormalization constants are defined as
\begin{align}\label{eq:nprCDef}
\mathcal{\hat O}_r( \mu) = \lim_{a\to 0} Z( \mu , a) \mathcal{\hat O}^{Latt}_b(a)\, .
\end{align}
Finally, the scheme is also converted to the $\overline{MS}$-scheme of~\cite{Buras:2000if} at the scale $\mu=3\ \rm{GeV}$.

As the operators $\mathcal{O}_{1+}^{++},\mathcal{O}_{1+}^{'++}$ and $\mathcal{O}_{2+}^{'++},\mathcal{O}_{2+}^{++}$ mix under renormalization, Z is a non-diagonal matrix.
To determine the non-zero elements, bare amputated vertex functions of the operators are computed on Landau gauge-fixed configurations and set equal to their tree level values in the chiral limit, $Z_{ik}\Lambda^{b}_{kj}/Z^2_q =  \Lambda^{tree}_{ij}/\Lambda^{tree}_{q}$.

We determined $Z$ on ensembles at three lattice spacings $a=0.09,0.12,0.15\ \rm{fm}$, and at several momenta and quark masses which allowed us to perform a chiral extrapolation and a momentum interpolation in order to determine $Z$ at the desired scale. 

\begin{figure}
\vspace{-0.5cm}
\includegraphics[scale=0.75]{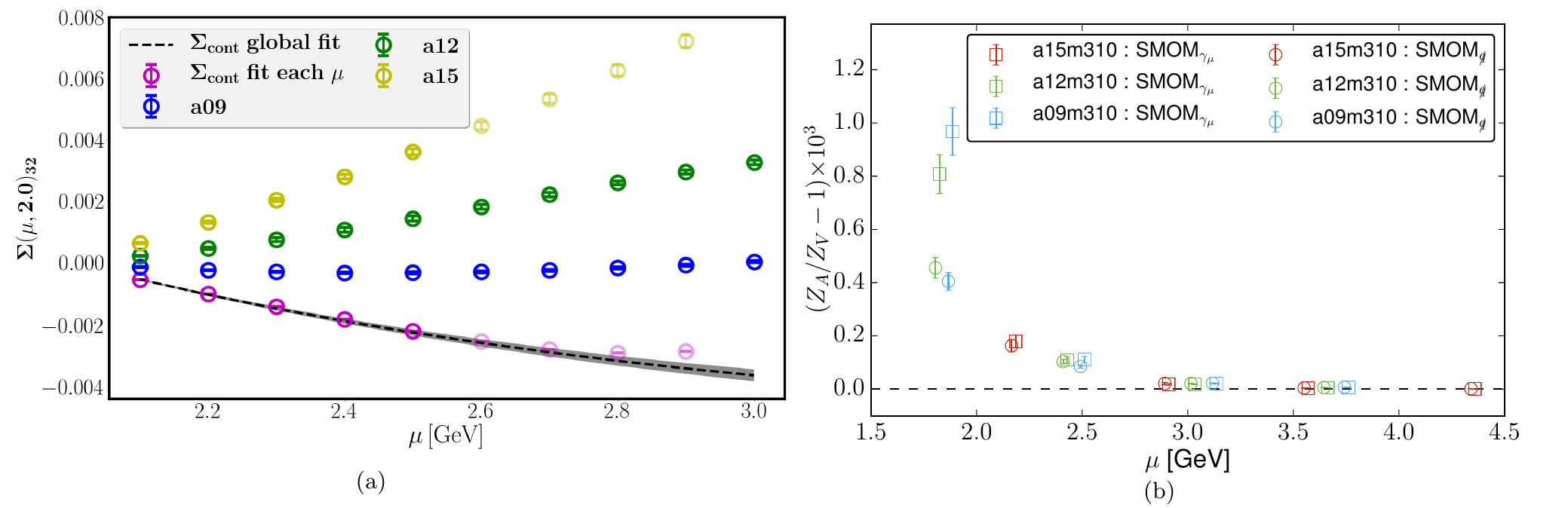} 
\caption{\label{sigmaFits}
a) Step scaling function fit.  The magenta points are fit at a fixed value of $\mu$ while the gray band is a fit to all $\mu$ simultaneously. b) $Z_A/Z_v$ ratio.}
\end{figure}

The chiral properties were checked by looking at the following quantities, $Z_A = Z_V $, $Z_P = Z_S$. This is illustrated in figure~\ref{sigmaFits}b where $Z_A/Z_V$ is shown to be very close to 1. Additionally, $Z$ entries for chirally forbidden matrix elements of the four-quark operators are observed to be orders of magnitude smaller than the allowed matrix elements. Also, off-diagonal $Z$ entries which mix the different operators are supressed with respect to the diagonal terms as these are subleading effects.

For the coarsest ensemble, $a=0.15\ \rm{fm}$, large errors were observed to appear below the desired scale, $\mu=3\ \rm{GeV}$. To obtain renormalization constants at the $\mu=3\ \rm{GeV}$ scale, running of the operator was employed to raise the renormalization scale. The technique that was adopted is known as step scaling and has been previously used together with the RI-SMOM scheme~\cite {Arthur:2010ht}. {The process consists in finding }the step scaling function which describes the running of the operators in the continuum limit. This step scaling function is defined by the "ratio" of two renormalization matrices at different momenta:
\begin{equation}
\Sigma(\mu_1,\mu_2,a) = Z(\mu_1,a)Z^{-1}(\mu_2,a)
\end{equation}
Furthermore, this quantity possesses a well defined continuum limit, $\Sigma(\mu_1,\mu_2)$, which we determined from a fit. For that purpose, the following fit anzats was employed:
\begin{equation}\label{eq:stepScalingFunc}
\Sigma(\mu_1,\mu_2,a) = \Sigma(\mu_1,\mu_2)+f(\mu)(b_2a^2+b_4a^4).
\end{equation}
In figure~\ref{sigmaFits}a, a sample plot shows the step scaling function data and fit. 
The renormalized matrix elements extrapolated to the physical point are given in Table II of Ref.~\cite{Nicholson:2018mwc}.
%

\bigskip\noindent\textbf{Four Quark Feynman-Hellmann Method:}
For the more complex two-nucleon contact interaction term, diagram~\ref{0nbbDiags}b) the calculation is far more expensive. To address this problem we are exploring an extension of the Feynman-Hellmann method developed in~\cite{Bouchard:2016heu} and used to compute $g_A$~\cite{Chang:2018uxx}, to compute 4-quark operators,
beginning with a test with a $\pi^-\rightarrow \pi^+$ matrix element.
Given an external source $S_\lambda=\lambda \int d^4x \bar{\psi}\Gamma^1\psi\bar{\psi} \Gamma^2 \psi$, a relation between the matrix element and two point functions can be determined (presented here for a pion with the leading wrap-around effect)
\begin{align}
\label{dmeff}
\frac{\partial m_{eff}}{\partial_\lambda}\biggr\vert_{\lambda=0}& =-\frac{\partial_\lambda C(t+\tau)+\partial_\lambda C(t-\tau)-2cosh(m_{eff}\tau) \partial_\lambda C(t)}{2\tau C(t)sinh(m_{eff}\tau)} 
= \frac{\bra{\pi}\mathcal{O}^{++}_{i+}\ket{\pi}}{2E_0^2} + e.s.,
\end{align}
with suppressed excited state (e.s.) contributions.
$\partial_\lambda C(t)$ contains the matrix element of interest:
{\begin{align}
\label{Nt}
N(t) =\int d^4x \left\langle \Omega \vert T{\mathcal{O}}(t)\bar{\psi}(x)\Gamma^1\psi(x)\bar{\psi}(x) \Gamma^2 \psi(x)\mathcal{O}^{\dagger}(0)\vert \Omega \right\rangle
\end{align}
To avoid calculating $N(t)$ directly (as it involves the operator insertion in the whole volume), instead a theory with the action $S=S_0+\int d^4x \frac{\sigma^2}{4}+\lambda_i' \int d^4x\ \bar{\psi}\Gamma^i \psi\ $, where $\sigma$ is a real scalar field, can be employed together with the Hubbard-Stratanovich Transformation~\cite{Stratonovich:1957jv,Hubbard:1959ub}.
This exchanges the calculation of four-quark operator insertions for bilinears, with the introduction of a new stochastic field.{ 
After integrating out the $\sigma$ field the four-quark operator is recovered in an analogous way to four-fermion interactions mediated by a W/Z boson, where the heavy W/Z bosons may be integrated out to give an effective interaction described by a four fermion contact term.}

\section{Conclusions}
We have presented the progress on the calculation of the short range interaction contributions to $0\nu\beta\beta$ from Lattice QCD. The leading contributions arising from the $\pi^- \rightarrow \pi^+$ process have been computed using  LQCD and have been renormalized to express the matrix elements in the {\MSbar}-scheme at $3\rm{GeV}$. 
The calculation of the $nn \rightarrow pp$ amplitude is much more challenging and numerically costly, though it will likely be necessary due to a potential breakdown of EFT power counting~\cite{Cirigliano:2018hja}.
We are therefore exploring variations of the Feynman-Hellmann method combined with a Hubbard-Stratonovich transformation to compute the corresponding matrix elements more efficiently.

\acknowledgments
The work of HMC was supported in part by the US DOE Nuclear Physics Double Beta Decay Topical Collaboration and the DOE Early Career Award Program.
Numerical calculations were performed with \texttt{Chroma}~\cite{Edwards:2004sx}, accelerated by QUDA~\cite{Clark:2009wm,Babich:2011np} and performed at LLNL through the LLNL Multiprogrammatic and Institutional Computing program through a Tier 1 Grand Challenge award, and on Titan, a resource of the Oak Ridge Leadership Computing Facility at the Oak Ridge National Laboratory, which is supported by the Office of Science of the U.S. Department of Energy under Contract No. DE-AC05-00OR22725, through a 2016 INCITE award.


\bibliographystyle{JHEP} 
\bibliography{Biblio} 

\end{document}